# V SIMPÓSIO ESTADUAL DE LASERS E APLICAÇÕES

26 a 28 de Outubro de 1992

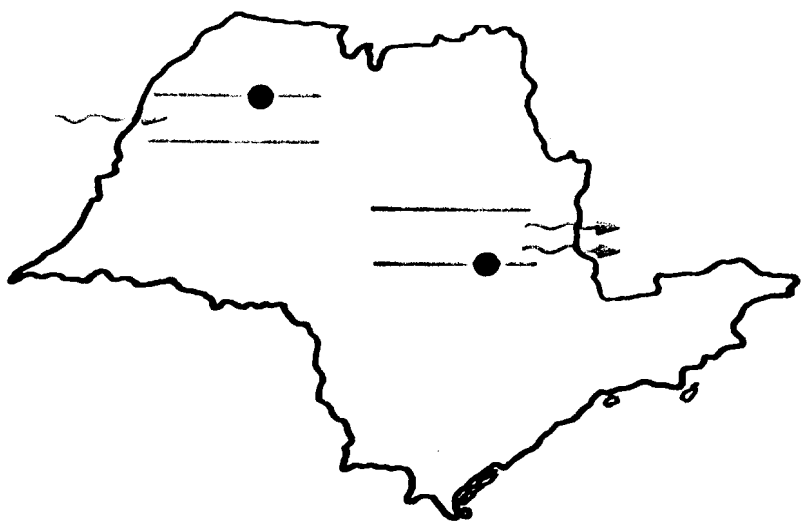

## Anais

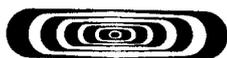

**ipen**

São Paulo - SP

# LENSLESS SLIDE PROJECTION: A DEMONSTRATIVE EXPERIENCE.

**Jose J. Lunazzi – Campinas State University**
bitnet: LUNAZZI at BRUC

*Abstract:*

*A 2D transparency may be projected on a diffractive screen by just illuminating it with a filament lamp of the same height. Sharpness of the filament width is naturally related to sharpness on the image, but some peculiar properties makes this experience to be different from a shadow projection case.*

INTRODUCTION

The development of many situations where we could obtain under white light certain imaging properties that could only be obtained before by means of laser light and holography, as for example the obtention of ortoscopic and pseudoscopic images in a continuous horizontal parallax [1] [2] [3] [4], lend us to the purpose of elliminating the lens that is needed in such experiences, as a way to better approach the classical holographic results. The following is an interesting result that, although do not render a 3D image, we hope that may be useful as an step toward lensless 3D diffractive imaging.

When a point luminous source illuminates a 2D transparency a sharp shadow may be seen on a common screen allowing for a simple case of lens-less projection. But point sources do not render bright shadows and if we use an extended filament lamp the shadow will no longer represent a clear image.
If we consider the filament to be in vertical position, a vertical blurring occurs at the shadow on the screen. If we directly look at the illuminated transparency we can only see the filament behind it being modulated by the particular portion of the transparency that lays along the visual line between our eyes and the filament.
If we use a diffraction grating of high spatial frequency as screen, we may clearly see the image of the transparency by looking at one diffraction order. It appears to be on the plane of the grating an in vertically multicoulored strips.

DESCRIPTION

The view of a horizontal line of the image is possible due to the diffraction dispersion at the grating as showed in **figure 1**, whose numbered elements are:

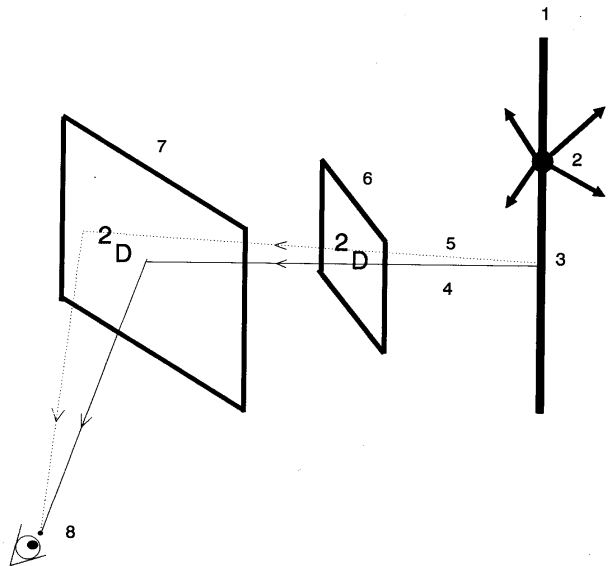

FIGURA 1

**1**, who represents the white light filament source from which each point (for example point **2**) is considered to send white light rays in any direction. Point **3** on the source is the one that coincides in height with the observer **8**, allowing us to select rays **4** and **5** that belong to the same horizontal plane. Wavelength of rays **4** is shortest than that os ray **5**.

Both rays traverse the black and white tranparency **6** (i.e. an slide) reaching the diffraction gratings **7** at two points from where they are diffracted to one eye of the observer.

The situation allows for considering those wavelengths that exactly satisfies the conditions for reaching the observer. We may then understand that if we generalize our case to the continuous case of any wavelength within the visible spectrum, we may see the transposition of the absorption information received by the beams as a horizontal line which has the colouring of a continuous spectrum. The caracters on the transparency, choosed as the symbols **2D** in our figure, may be seen clearly without needing to precisely adjust any distance between the elements, as should be in the case of focusing with a lens.

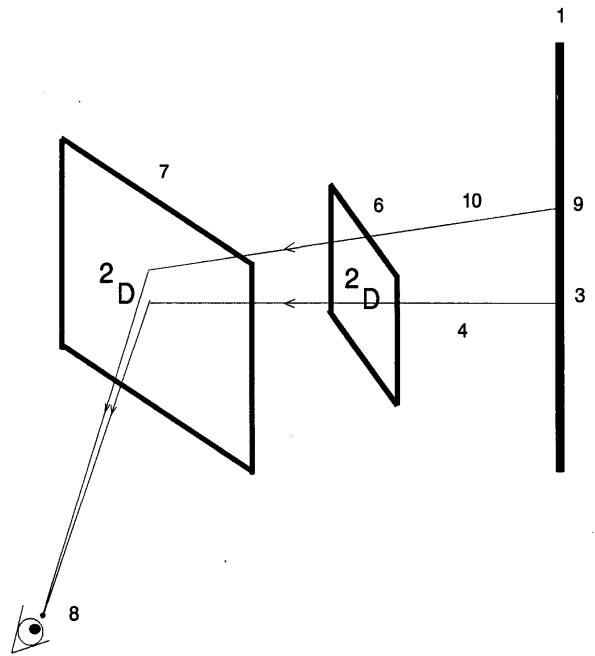

FIGURA 2

In **figure 2** we explain how the vertical dimension may also be viewed by considering a different point **9** in the souce which is at a different vertical position from our previous point **3**. We consider for our purpose a ray **10** whose wavelength is the same of the one we considered first for ray **4**. Diffraction conditions indicates that this ray may reach the observer giving to him vertical information on the figure.
So that the horizontal field of view is directly related to the diffraction capacity of the grating and the vertical field depends on the heigth of the source. Horizontal resolution certainly depends on the width of the source while vertical resolution seems to be only affected by the correct alignment between the filament and the grating.

EXPERIMENTAL DETAILS

We performed the experience by using a common domestic filament lamp of 110 VCA and 25 W whose filament was 65 mm in height and about 0,2 mm wide.

We put it in a vertical position illuminating a black and white 35 mm slide and at 14 cm from it the light reached a common plastic holographic transmission grating of 590 lines/mm, whoses lines were vertically aligned.

It was then easy to see the image of the transparency on the grating, with a little degree of horizontal amplification, but whose quality was enough for visual observation. It should be impossible to recognize any of the characters written on the transparency if instead of a diffraction grating we would use a common white screen.

The degree of freedom for the diffractive screen distance was as high as to allow for doubling its initial distance without loosing resolution, just perturbed by the doubling of the lateral amplification.

Observation at the first and second diffraction orders showed the extension of the horizontal field for the second case.